# Information Security Policy: A Management Practice Perspective

**Moneer Alshaikh**
Department of Computing and Information Systems
Melbourne School of Engineering
University of Melbourne
Victoria, Australia
Email: Malshaikh@student.unimelb.edu.au

**Sean B. Maynard**
Department of Computing and Information Systems
Melbourne School of Engineering
University of Melbourne
Victoria, Australia
Email: Sean.Maynard@unimelb.edu.au

**Atif Ahmad**
Department of Computing and Information Systems
Melbourne School of Engineering
University of Melbourne
Victoria, Australia
Email: Atif@unimelb.edu.au

**Shanton Chang**
Department of Computing and Information Systems
Melbourne School of Engineering
University of Melbourne
Victoria, Australia
Email: Shanton.chang@unimelb.edu.au

## Abstract

Considerable research effort has been devoted to the study of Policy in the domain of Information Security Management (ISM). However, our review of ISM literature identified four key deficiencies that reduce the utility of the guidance to organisations implementing *policy management practices*. This paper provides a comprehensive overview of the management practices of information security policy and develops a practice-based model. The model provides comprehensive guidance to practitioners on the activities security managers must undertake for security policy development and allows practitioners to benchmark their current practice with the models suggested best practice. The model contributes to theory by mapping existing information security policy research in terms of the defined management practices.

**Keywords**: Information security policy, Policy development, Security policy management practice

## 1 Introduction

There is growing recognition of the role of management in protecting organisational information from a range of security risks such as: leakage of trade secrets and intellectual property, disruption of mission-critical systems, and malicious attack from both insiders and outsiders (Ahmad et al. 2014a; Alshaikh et al. 2014; Webb et al. 2014) Policy is a critical formal control by which senior management provides strategic and tactical guidance on a range of issues such as what security structures, roles, and processes must be instituted and the acceptable use of information technologies (Ahmad et al. 2014b; Sommestad et al. 2014). Consequently, security researchers have consistently argued that the effectiveness of managerial practices associated with security policy is critical to a successful security program (Maynard and Ruighaver 2006; Siponen et al. 2014).

Our review of both professional and academic literature reveals that considerable research effort and progress has been made on the provision of high-level policy management lifecycles or models for organisations. However, there are a number of deficiencies that reduce their utility to organisations seeking guidance on what managerial practices are involved in implementing security policy. The literature: lacks a holistic view of the policy lifecycle (deficiency 1); lacks consistency in terminology and semantics (deficiency 2); uses varying levels of granularity in describing policy management activities (deficiency 3); and makes it difficult to extricate guidance on policy management from that of other practice areas such as risk management and Security Education, Training, and Awareness (SETA) (deficiency 4).

Therefore, the aim this paper is to: (1) provide a comprehensive overview of the management practices of information security policy; and (2) develop a practice-based model that addresses the four aforementioned deficiencies. The study addresses the following research question:

*What information security policy management practices should be implemented in organisations?*

This paper is organised as follows. First, we review existing policy management lifecycles in the background section. Second, we explain the research methodology employed to review and analyse the literature. Third, we propose a model of managerial practices related to security policy. Fourth, we explain how the proposed model addresses the identified deficiencies in the discussion section. Finally, we revisit the main contribution and conclude with implications of the research.

## 2  Background

There are a number of studies on the development and implementation of information security policy (Bayuk 1997; Kadam 2007; Knapp et al. 2009; Rees et al. 2003; SANS Institute 2001; Whitman 2008). The majority of these studies present the development of security policy as multi-stage lifecycles. Using a lifecycle approach to develop security policy is very beneficial as it allows good management of the process of security policy development and assures managers that all important activities for the development process are performed (Maynard and Ruighaver 2007; Patrick 2002). Ølnes (1994) stresses the importance of having a methodological approach in developing, implementing and maintaining security policy. Further, Patrick (2002) argues that the use of a security policy lifecycle approach will ensure a comprehensive development process that encompasses all required activities to develop an effective security policy.

Previously four deficiencies have been identified in existing policy development lifecycles. In this section these deficiencies will be discussed in detail. Table 1 summaries deficiencies of existing policy development models.

The first deficiency is the lack of holistic view of the policy lifecycle. This can be identified clearly in some of the existing policy development lifecycles. For example, Bayuk (1997) presents a process with a narrow view that focuses on the development of policy documents and does not include any practices related to the implementation and the maintenance of the policy. Bayuk (1997)'s process consists of several steps. It starts by identifying assets and then forming a team to develop the policy. Then the draft policy is produced. The draft policy goes through a review process leading to approval and publishing. Researchers (e.g. Patrick (2002) suggest that the development of security policy goes beyond the development of the document. Similar to Bayuk (1997), Ølnes (1994) model of policy development is not holistic in that it does not specifically address how policy document is developed, communicated, enforced and evaluated. A recent paper by Al-Mayahi and Sa'ad (2014) focuses on developing a detailed information security policy, rather than providing guidance on the development process of the policy.

The second deficiency is that existing policy development lifecycles lack consistency in terminology and semantics. While (Hare 2002; Karyda et al. 2005; Lowery 2002; Patrick 2002; Whitman 2008) present a more holistic view of the policy development process, there are few overlapping concepts such as compliance, monitoring and enforcement. These three concepts are presented in the approach as three distinct activities, while they represent the management efforts to ensure that the policy is being adhered to by employees. Referring to one concept in three different terms or referring to different activities in one term may cause confusion among security practitioners embarking on the process of policy development.

The third deficiency that has been identified is that existing policy development lifecycles use varying levels of granularity in describing policy management activities. Each of the policy lifecycles differs in the level of detail and emphasis on policy development aspects. For example, Hare (2002) presents the development process of security policy in a systematic way, however, details are lacking about how the policy will be published (what form it will take e.g. online, HTML) and how it will be communicated and enforced. In addition, Hare (2002) did not discuss the issue of user compliance with the policy and the importance of user awareness and training in communicating and enforcing security policy in organisations. The problem with the depth of content can also be seen in the policy development lifecycles proposed (Bin Muhaya 2010; Klaic and Hadjina 2011; Knapp et al. 2009; Lowery 2002; Whitman 2008; Whitman et al. 1999; Wood 1995). The authors provide scant detail about many important activities in the development process of security policy. For example, the lifecycle developed by Whitman (2008) does not provide guidance on communicating and enforcing the policy. Further, Knapp et al. (2009) proposes a model of policy development that presents the process in a very general manner without providing sufficient descriptions of the policy management practices.

| Policy development | Deficiency 1 | Deficiency 2 | Deficiency 3 | Deficiency 4 |
|---|---|---|---|---|
| Rees et al. (2003) |  | X |  | X |
| Patrick (2002) |  | X |  | X |
| Knapp et al. (2009) |  |  | X | X |
| Karyda et al. (2005) |  |  | X |  |
| Kadam (2007) | X |  | X |  |
| Hare (2002) |  |  | X | X |
| Bayuk (1997) | X | X |  |  |
| Wood (1995) |  |  | X | X |
| Ølnes (1994) | X |  |  | X |
| Whitman et al. (1999) |  |  | X |  |
| Saltzman and Gadkari (2004) | X |  | X |  |
| Whitman and Mattord (2008) |  |  | X |  |
| Lowery (2002) |  |  | X | X |
| Al-Mayahi and Sa'ad (2014) |  |  | X | X |

*Table 1 Summary of Deficiencies identifies in Existing policy development lifecycles*

The fourth deficiency is the difficulty to extricate guidance on policy management from that of other practice areas such as risk management and SETA. This is because models such as that proposed by Ølnes (1994), Rees et al. (2003), Knapp et al. (2009) and Patrick (2002) include practices such as conducting risk assessment, development of security awareness program and selection of technical controls as part of policy development lifecycle. We acknowledge the importance of having risk assessment as an input the policy development process, as well as the need for security awareness and training to communicate and enforce policy. However, we argue that conducting risk assessment and developing a security awareness and training program are not part of the security policy lifecycle. In fact, policy development lifecycles proposed by Ølnes (1994) and Rees et al. (2003) go beyond the development of security policy to the development of a security program in the organisation. They address security policy, risk assessment, technical controls and incident response. Security policy is only part of the overall security program that the model focuses on.

Our review of the literature shows evidence of four critical deficiencies that affect organisations seeking to implement security policy. The review also supports Knapp et al. (2009)'s assertion that there is a need for empirical research in the area as the majority of existing policy development lifecycles are conceptual and lacking support from empirical data. Therefore, there is a need to develop a comprehensive, coherent and empirically tested security policy management practices model that addresses the four deficiencies in literature.

## 3   Research Approach

We conducted a comprehensive and rigorous review of the information security policy literature. For both academic and professional literature, we used the following keywords to search SpringerLink, IEEE Xplore, ScienceDirect, the ACM digital library, ProQuest and Google Scholar: 'information security policy', 'information security policy development', 'security policy management', 'policy development lifecycle'.

The preliminary results consisted of 132 scholarly articles, industry standards, and technical reports. A review of abstracts resulted in the elimination of 20 papers that were not related to security policy, leaving 112 security policy related papers dated between 1994 and 2015. Twenty publications were related to the development process of security policy, fourteen articles (journal and book sections) propose security policy lifecycles and 92 publications discuss specific aspects of security policy such as policy quality, compliance and employees' attitude towards security policy.

A coding process was utilised to synthesise the identified articles to develop a comprehensive and robust understanding of security policy management. Furthermore, a security policy management model was proposed based on the understanding emerged from the review and synthesis process.

The guidelines proposed by Okoli and Schabram (2010) were followed to review and analyse the literature. The review process focused first on the fourteen articles proposing security lifecycles. Each article was reviewed; paragraphs reduced into themes, and sentences that related to policy development were underlined. Then ideas and concepts were recorded on the margins. Once the entire paper was reviewed, the important concepts were summarised in the back of its last page. The summaries enable the researcher to remember the important themes discussed throughout the paper by the end of the overall review. After the fourteen policy development lifecycle articles were reviewed, the coding process was used to synthesise the articles. The coding process includes open, axial and selective coding as described in Neuman (2006).

The second review started with more focus on the underlined excerpts, and summaries resulting from the first review. Themes related to policy management began to emerge. The researchers reviewed the identified themes giving more attention to themes that are frequently discussed throughout the articles. Themes were divided into sub-themes, and several related concepts were combined into more general one. A comparison was made between the themes that reappear in different places.

A similar process was applied to review the 92 publications that do not directly address the development process of security policy. However, the review process was informed by the result of the review of security policy lifecycle. No new themes were identified, however, the review provided more details about the identified themes from the lifecycles, and located evidence to support the identified themes. For example, some security policy lifecycles mention the importance of involving stakeholders in the development process of security policy. However, they did not identify who the stakeholders were nor did they discuss their roles and responsibilities in the policy development process. These details could be identified in some of the 92 additional papers.

The review process was guided by our definition of security policy management practices. We define policy management practices as the strategic-level activities undertaken to manage security policy in organisations. Managing security policy involves the development, implementation and evaluation of security policy.

The coding process eventually led to the identification of seven security policy management practices. Each practice has several activities. These practices are grouped into three stages.

## 4   Information Security Policy Management Practice Model

The overall understanding that emerged from the systematic review and coding process resulted in the development of a model of information security policy management practices. This section discusses the proposed model. Table 2 depicts the model which consists of three main stages: the *Development* stage, the *Implementation & Maintenance* stage, and the *Evaluation* stage. Each stage consists of a number practices, each having a number of activities.

| Stage | Practice | Activities |
|---|---|---|
| Develop | Establish information security policy development team | Identifying key stakeholders |
| | | Define roles and responsibilities |
| | Determine the security needs of the organisation | Identify security requirements |
| | | Assessing the organisation's current policies and procedures |
| | Compiling security policy document | Select policy components |
| | | Draft security policy |
| | | Review draft policy document |
| Implement & Maintain | Distribute policy | Select policy delivery methods |
| | | Doing the actual distribution |
| | Communicate policy | Communicate policy through various ways such as briefing, seminars, and awareness campaign. |
| | Enforce policy | Undertake various activities to enforce policy such as implementing technological mechanisms and conducting SETA program |
| Evaluate | Periodically review information security policy | Collect feedback from relevant stakeholders about security policy |
| | | Examine security incidents' reports and new risk assessment |

*Table 2. Information Security Policy Management Practices Model*

## 4.1 The Development stage

The development stage of the process of managing information security policy represents all the practices associated with the development of security policy.

### 4.1.1 Establish information security policy development team

The first practice that information security managers in organisations must undertake in the process of developing information security policy (ISP) is to establish the policy development team. There are two main activities in this practice: (1) identify key stakeholders who should be involved the development of policy and (2) define roles and responsibilities.

- Identifying key stakeholders

The involvement of relevant stakeholders in the security policy development process is a success factor for security policy in the stages of development, implementation and evaluation. Therefore, a team of representative stakeholders from across the organisation at all levels is assembled. Representative stakeholders in the organisation may include technical personnel, process owners, decision makers, managers, legal department, the human resource department, users, plus other function area personnel affected by the new policy (Maynard et al. 2011; Ølnes 1994; Wood 1995). The scope of the developed policy is an important factor to determine who should involve in the development process (Patrick 2002). For example, a security policy developed for a specific department within the organisation may involve less people in the development process than the policy developed for the entire organisation.

- Define roles and responsibilities

It is important to clearly define the roles and responsibilities of development team members to avoid delays in the development process due to interpersonal challenges and political objections that may occur (SANS Institute 2001; Whitman and Mattord 2010; Wood 1995). Maynard (2010) asserts that while many authors emphasize the importance of involving different stakeholders in the development process; the roles of these stakeholders remain unclear. He also points out that authors simply mention the name of the stakeholder that needs to be involved in the development process without specifying what this group of people should do in the process. Therefore, Maynard (2010) discusses the roles of each stakeholder in the development process of security policy.

### 4.1.2 Determine the security needs of the organisation

After establishing the policy development team, the organisation should determine its security needs (Rees et al. 2003; SANS Institute 2001; Whitman and Mattord 2010). A good understanding of the current situation of the organisation, as well as sufficient understanding of the organisation's security goals and objectives is required (Ølnes 1994; Palmer et al. 2001; Stahl et al. 2012). This can be done by conducting a thorough investigation of the problem facing the organisation (Whitman 2008). Determining the security needs of the organisation consists of two activities: (1) identify security requirements and (2) assessing the organisation's current policies and procedures

- Identify security requirements

Due to the fact that organisations have different security needs, organisations have different security requirements and objectives (Karyda et al. 2005; Ølnes 1994; Wood 1995). Baskerville and Siponen (2002) argue that it is important to have a good understanding of the organisation's security requirements when developing security policies. Therefore, the organisation should identify its security requirements, including the level of security that the organisation aims to achieve. Security requirements should specify the requirements of the organisation for addressing security risks, identified through risk assessment, in order fulfil its security needs and achieve its business objectives.

The result of the risk assessment is an input to identify security requirements, therefore, some authors include risk assessment as a practice in their security policy lifecycles (Bayuk 1997; Gaunt 1998; Rees et al. 2003). However, although the result of risk assessment is a prerequisite to identify the security requirements, assessing risk should be part of security risk management, not policy development.

- Assessing the organisation's current policies and procedures

Assessing currently implemented security policy and procedures has several benefits. First, it aids the security development team in understanding the current status of existing policy and procedures (Doherty and Fulford 2006; Palmer et al. 2001; Rees et al. 2003; Whitman 2008). This is important as it allows the organisation to identify gaps in current policy and to determine whether the existing policy will help the organisation to address risk, by meeting its security requirements, therefore identifying areas that need to be addressed by the new policy. Second, assessing existing policies and procedures will ensure that new policies conform to existing policy standards (SANS Institute 2001). This will increase the chance of successful implementation of the updated policies in the organisation (Peltier 2013). Third, the assessment process helps gather key materials such as existing policy and procedures documents, which will be used by the development team as key reference (Patrick 2002; Whitman et al. 2001).

### 4.1.3 Compiling the security policy document

Compiling the security policy document is the last practice in the development phase of information security policy. The security policy document should state the management commitment and direction, and set out the organisation's approach to manage information security (ISO/IEC27002 2006). Maynard and Ruighaver (2003) argue the importance of documenting the information security policy development process to justify the development process itself and also to aid in the evaluation of existing policy.

Compiling the security policy document consists of a number of activities, including: selecting policy components, writing draft policy and presenting the draft policy to relevant stakeholders for review, comment and then approval (Hare 2002; Patrick 2002; Whitman 2008).

- Select policy components

The policy development team selects policy items to address the security needs of the organisation (Lowery 2002; Rees et al. 2003; Wood 2005). Policy items may address access control, Internet usage, the use of mobile devices and portable storage devices and so forth. For example; access controls' items should discuss authorised access to the systems, ways to control access (passwords and/or biometrics) and consequences of unauthorised access (Whitman et al. 1999; Wood 2005).

- Draft security policy

The policy development team should appoint one of its members to write the policy (Anderson Consulting 2000). The rest of the team should provide guidance on context and the content of the policy. Höne and Eloff (2002b) explore the factors that make security policy an effective control in protecting organisational information assets. They report on characteristics that should be considered when writing security policy. These characteristics are concerned with the length and writing style. A security policy document should be short because if it is too long, the users will not read it. Many authors (e.g. Stahl et al. 2012; Whitman 2008; Wood 1995) highlight the importance of using an appropriate language in writing security policy. They suggest that security policy should be written in a clear, concise, and easy-to- understand language.

- Review draft policy document

Once the first draft of the policy is created, it should be presented to relevant stakeholders to review and provide feedback about quality, usability and acceptance of the policy (Kadam 2007; Lindup 1995; Whitman 2008). Feedback on policy should be sent to the author to update the policy. Policy writing and revision are an iterative process (Rees et al. 2003). In other words, the draft may progress through many revisions until the final policy is produced. The final policy will be sent to top management for final approval. Then it will be published be ready to be implemented (Whitman 2008).

## 4.2 The Implementation and Maintenance Stage

The implementation and maintenance stage is the second stage of the security policy management process. It is an ongoing process, which consists of several practices. Following is a discussion of information security management practices within this stage.

### 4.2.1 Distribute policy

The practice of distributing the policy is to ensure that all stakeholders in the organisation, including users and mangers, have access to the policy document (Höne and Eloff 2002a). Effective dissemination of the policy to the individual affected by the policy requires a substantial effort from organisation in order to be done effectively (Whitman 2008). The distribution of the policy involves: (1) Selecting the delivery methods and (2) using the delivery methods to deliver the policy.

- Select policy delivery methods

There are various ways to distribute the policy in the organisation (Gaunt 1998; Lindup 1995; SANS Institute 2001; Whitman 2008). While some organisations prefer a hardcopy dissemination in which a printed copy of the document is delivered to the employees, others publish the policy electronically through email and internal and external network (Whitman 2008). No matter what methods the organisation chooses to distribute the policy, it should be available and easy to access (SANS Institute 2001). Therefore, the organisation should select the most appropriate policy delivery methods to ensure that the policy reaches the people it is applied to. The selection of the delivery methods depends on the organisation environment and the preference of the employees.

- Doing the actual distribution

After the selection of the delivery methods, the policy should be prepared in the appropriate format, whether it HTML, PDF or a Word document (Anderson Consulting 2000; Hare 2002). The format is guided by the delivery methods selected and the organisations preferences. Once the appropriate format is prepared the distribution of the policy takes place.

### 4.2.2 Communicate policy

By distributing the policy, the organisation has no guarantee that individuals who received the policy will actually read it. Therefore, the organisation must communicate the policy (Rees et al. 2003; SANS Institute 2001; Sommestad et al. 2014). Communicating the policy is an essential practice before the enforcement of the policy (Knapp and Ferrante 2012; Siponen et al. 2007; Whitman et al. 2001). Successful communication of the policy leads to better compliance from employees (Sommestad et al. 2014).

Communicating the policy is important in assisting the organisation to manage changes in its processes caused by the new policies implementation (Maynard and Ruighaver 2003). Communicating the policy has three main objectives: to make users aware of the policy, to communicate reason for implementing the policy, making users aware of how it will affect them and what the implications are if they do not comply (Knapp et al. 2009; Maynard and Ruighaver 2003).

There are a number of ways to communicate the policy, including using security education, training and awareness (SETA) programs. Siponen et al. (2014) emphasises the importance of a SETA program in teaching the organisation's employees about their role in maintaining the policy so that policy becomes "as an integral part of their job". In the same vein, Whitman (2008) stresses the significant role that an awareness program plays in keeping policies fresh in employees' minds.

For example, communicating the policy is done through conducting training sessions to teach users how perform security procedures that the policy requires. Another example is to use an awareness campaign to raise people's awareness about the organisation's policy. Policy also can be communicated through a monthly briefing to ensure that employees not just understand the policy, but also have the necessary skills to adhere to the policy guidelines.

### 4.2.3　Enforce policy

Enforcing policy is an ongoing activity to ensure that the policy is adhered to (Hare 2002; Lowery 2002). Policy enforcement does not simply involve identifying and penalizing violators. Enforcement is a managerial activity that considers the unauthorized act itself, as well as the severity of the offence and user's intent (Puhakainen and Siponen 2010).

The literature emphasises the importance of policy enforcement and that without enforcement the security policy has no value (Doherty and Fulford 2006; Knapp et al. 2009; Rees et al. 2003; Whitman 2008). The SANS Institute (2001) reports that to mitigate risks to information security the "policy must be enforced in a strict manner, and noncompliance must be punished" *(p8)*. Enforcement and compliance needs to be in place to ensure effective implementation of security policy (Al-Mayahi and Sa'ad 2014).

In order to enforce policy a number of activities need to be accomplished. First, the implementation of technological mechanisms such as user administration (adding, deleting and modifying system and application users), evaluation and applying security patches to systems and applications, system and application monitoring for security events and administering anti-virus applications (Li et al. 2014; Rees et al. 2003). Second, enforcement can be done through conducting a SETA program to change employees' behaviour towards adherence to security policies (Siponen et al. 2014; Sommestad et al. 2014).  Sommestad et al. (2014), Li et al. (2014) and (Vance et al. 2012) argue that organisations should shift from enforcing policy through the implementation of incentives and sanctions, towards creating a shared vision of security policy. This argument supports the claim made by several authors (e.g. Hassan and Ismail 2012; Lim et al. 2010; Oost and Chew 2012; Ramachandran et al. 2012; Ruighaver et al. 2007) that establishing security culture will result in better compliance with security policy.

## 4.3　The Evaluation Stage

An effective security policy requires constant review and revision. The Evaluation stage has two main objectives: (1) to determine if the policy still effective and (2) to identify the needs to update policy to incorporate to organisational changes. The process of evaluating the security policy serves as a feedback mechanism providing input for the development of the policy.

### 4.3.1　Periodically review information security policy

There is wide agreement in the literature that policy needs to be reviewed periodically (Knapp et al. 2009; Maynard and Ruighaver 2003; Rees et al. 2003). The organisational environment, both internal and external, changes constantly. This leads to changes in the information risks faced by the organisation. In order for the information security policy to continue to be current, effective and relevant, the policy needs to be modified. To accomplish the review practice two main activities should be carried out.

- Collect feedback from relevant stakeholders about security policy

Feedback can be collected from relevant stakeholders (managers, users …etc.) using interviews and surveys and other data collection means (Anderson Consulting 2000). The feedback should be analysed to determine the effectiveness of the policy, to monitor compliance and to determine the relevance of the policy. This will help to identify whether the organisation needs to modify the policy and helps to avoid the risk of having an outdated and irrelevant security policy, thus being an ineffective control in mitigating risks (Anderson Consulting 2000; Patrick 2002).

- Examine security incidents' reports and new risk assessment

The importance of gathering security incident data to inform policy development cannot be underestimated. The number and type of incidents can be strong indicators to determine where the policy is no longer effective (Bañares-Alcántara 2010; Kadam 2007; SANS Institute 2001). This helps to identify areas in the existing policy that must be updated, added, or removed. In other words, it helps to recommend possible changes in the current policy to ensure that the organisation's security policy remains an effective control in protecting the organisation from the evolving risks' environment.

In terms of when a policy review takes place, several researchers suggest that the review and revision of the policy should be done at least annually (Höne and Eloff 2002a). Others, however suggest that it should occur whenever major changes in information systems of the organisation are made (Palmer et al. 2001; Sommestad et al. 2014; Wood and Lineman 2009). Security incidents may also trigger the process as well (Ahmad et al. 2015; Park et al. 2012).

The management of information security policy is an iterative process. Therefore, the review practice provides a valuable feedback on the current policy (the need to change and update the policy) to the development stage in the policy management practices.

# 5 Discussion

In order to address the deficiencies identified in the literature, the proposed model of security policy management practices (see Table 1) offers the following four contributions.

## 5.1 Provide a more holistic view of the policy management process

Patrick (2002) argues that organisations should take a more holistic view of the policy development process than the simple writing and implementation tasks. Further he added that taking a narrow view of the process results in "developing policies that are poorly thought out, incomplete, redundant, not fully supported by users or management, superfluous, or irrelevant" (p297). Therefore, to ensure that the proposed model provides a more holistic view of the policy management process, a comprehensive and systematic review of security policy related literature has been conducted. Qualitative analysis techniques including coding and discussion were utilized to construct holistic view of the process.

## 5.2 Improve the consistence in terminology and semantics

The proposed model of security policy management practices addresses the problem of inconsistency in terminology and semantics by presenting a clear understating of the terminology that is employed to refer to policy management activities. For example, the model makes a clear distinction between 'communicating' and 'distributing' the security policy, which has been interchangeably used in the literature. The proposed model refers to selecting the policy delivery methods and doing the actual delivery of the policy documents to the employees as 'distribute the policy' while ensuring that the policy has been read and understood by employees is referred to as 'communicate the policy'. Another example of inconsistency in terminology and semantics is the use of 'enforcement' and 'compliance' to refer to the effort that management should do to ensure that the policy is adhered to. The model defines the management practice to ensure that users adhere to policy as 'enforce policy'. Compliance, in the other hand is the desired result of the enforcement practice.

## 5.3 Discuss policy management activities at the appropriate level of granularity

The proposed model focuses on the management practices to manage security policy. The model is organised in three institutionalisation stages. Each stage consists of several management practices, and each practice consists of activities should be undertaken to perform this practice. This organisation of the model provides in depth discussion of the management practices of security policy, which enable sufficient guidance for the organisations to manage their security policy.

## 5.4 Simplify guidance on policy management

In order to address the difficulty to extricate guidance on policy management from that of other practice areas such as Risk and SETA, the proposed model focuses purely on policy management practices. The omission of practices of other areas of security management does not mean that the proposed model ignores these practices (these will be discussed in future work) and their importance in the process of managing security policy, but rather it aims to simplify guidance on policy management. For instance, while the model does not consider conducting risk assessment as a policy management practice, it acknowledges the importance of having resent risk assessment report during the development as well as the evaluation stage of policy management process. Further, the model reports the need for conducting SETA program to communicate and enforce security policy, however, the development and implementation of SETA program is not part of the policy management process.

## 6   Conclusion & Future Research

This paper has discussed the development of a model of security policy management practices. The review and analysis of the literature has provided a more comprehensive and rigorous understanding of the security policy development process. From this review, we have developed a model of information security policy management practice. The model consists of three institutionalisation stages: the development stage, the implementation and maintenance stage, and the evaluation stage. Each stage consists of several practices containing management activities.

The security policy management practices model has several implications for practitioners and researchers. The model will provide comprehensive guidance on security policy management practices that can be implemented to manage security policy in organisations. The model will also allow practitioners to benchmark their security policy management activities against the model and provide a better understanding of the process.

The model will allow researchers to map existing policy management research activity to the proposed model (i.e. institutionalisation stages as well as practices within each stage) to identify areas for future research.

The model provides a sound basis for further work. The next step is to empirically refine and validate the model using, in turn, a set of expert interviews, a set of case studies within Australian organisations and finally a set of focus groups. The expert interviews will be conducted to gain comment on the proposed model for the purpose of refinement. The case studies will allow the assessment of security management practices implementation against the model. Finally, the focus groups will perform the final validation of the model.

## 7   References

Ahmad, A., Bosua, R., and Scheepers, R. 2014a. "Protecting Organizational Competitive Advantage: A Knowledge Leakage Perspective," *Computers & Security* (42), pp 27-39.

Ahmad, A., Maynard, S.B., and Park, S. 2014b. "Information Security Strategies: Towards an Organizational Multi-Strategy Perspective," *Journal of Intelligent Manufacturing* (25), pp 357-370.

Ahmad, A., Maynard, S.B., and Shanks, G. 2015. "A Case Analysis of Information Systems and Security Incident Responses," *International Journal of Information Management*).

Al-Mayahi, I.H., and Sa'ad, P.M. 2014. "Information Security Policy Development," *Journal of Advanced Management Science* (2:2), June , pp 135-139.

Alshaikh, M., Ahmad, A., Maynard, S.B., and Chang, S. 2014. "Towards a Taxonomy of Information Security Management Practices in Organisations," *25th Australasian Conference on Information Systems*, Auckland, New Zealand.

Anderson Consulting. 2000. "Policy Framework for Interperting Risk in Ecommerce Security," Center for Education and Research in Information Assurance and Security, Purdue University.

Bañares-Alcántara, R. 2010. "Perspectives on the Potential Roles of Engineers in the Formulation, Implementation and Enforcement of Policies," *Computers & Chemical Engineering* (34:3), March , pp 267-276.

Baskerville, R., and Siponen, M. 2002. "An Information Security Meta-Policy for Emergent Organizations," *Logistics Information Management* (15:5/6), pp 337-346.

Bayuk, J. 1997. "Security Through Process Management," *Morristown, NJ, Price Waterhouse*.

Bin Muhaya, F.T. 2010. "An Approach for the Development of National Information Security Policies," *International Journal of Advanced Science & Technology* (21), August , pp 1-9.

Doherty, N.F., and Fulford, H. 2006. "Aligning the Information Security Policy with the Strategic Information Systems Plan," *Computers & Security* (25:1), September , pp 55-63.

Gaunt, N. 1998. "Installing an Appropriate Information Security Policy," *International Journal of Medical Informatics* (49:1), pp 131-134.

Hare, C. 2002. "Policy Development," in: *Information Security Management Handbook Fourth Edition, Volume 3*. CRC Press, pp 353-383.

Hassan, N.H., and Ismail, Z. 2012. "A Conceptual Model for Investigating Factors Influencing Information Security Culture in Healthcare Environment," *Procedia - Social and Behavioral Sciences* (65:0), pp 1007-1012.

Höne, K., and Eloff, J.H.P. 2002a. "Information Security Policy — What Do International Information Security Standards Say?," *Computers & Security* (21:5), October , pp 402-409.

Höne, K., and Eloff, J.H.P. 2002b. "What Makes an Effective Information Security Policy?," *Network Security* (2002:6), pp 14-16.

ISO/IEC27002. 2006. "Australian/New Zealand Standard: Information Technology - Security Techniques- Code of Practice for Information Security Management."

Kadam, A.W. 2007. "Information Security Policy Development and Implementation," *Information Systems Security* (16:5), pp 246-256.

Karyda, M., Kiountouzis, E., and Kokolakis, S. 2005. "Information Systems Security Policies: A Contextual Perspective," *Computers & Security* (24:3), pp 246-260.

Klaic, A., and Hadjina, N. 2011. "Methods and Tools for the Development of Information Security Policy - a Comparative Literature Review," *MIPRO, 2011 Proceedings of the 34th International Convention*, pp. 1532-1537.

Knapp, K.J., and Ferrante, C.J. 2012. "Policy Awareness, Enforcement and Maintenance: Critical to Information Security Effectiveness in Organizations," *Journal of Management Policy and Practice* (13:5), Dec 2012, pp 66-80.

Knapp, K.J., Franklin Morris Jr, R., Marshall, T.E., and Byrd, T.A. 2009. "Information Security Policy: An Organizational-Level Process Model," *Computers & Security* (28:7), 10//, pp 493-508.

Li, H., Sarathy, R., Zhang, J., and Luo, X. 2014. "Exploring the Effects of Organizational Justice, Personal Ethics and Sanction on Internet Use Policy Compliance," *Information Systems Journal* (24:6), pp 479-502.

Lim, Ahmad, A., Chang, S., and Maynard, S. 2010. "Embedding Information Security Culture Emerging Concerns and Challenges," PACIS 2010 Proceedings, paper 43 , pp 463-474.

Lindup, K.R. 1995. "A New Model for Information Security Policies," *Computers & Security* (14:8), pp 691-695.

Lowery, J. 2002. "Developing Effective Security Policies." Dell power solutions.

Maynard, S., and Ruighaver, A. 2003. "Development and Evaluation of Information System Security Policies," *Information Systems: The Challenges of Theory and Practice*, pp 366-393.

Maynard, S., and Ruighaver, A. 2006. "What Makes a Good Information Security Policy: A Preliminary Framework for Evaluating Security Policy Quality," *Proceedings of the fifth annual security conference, Las Vegas, Nevada USA*, pp. 19-20.

Maynard, S., and Ruighaver, A. 2007. "Security Policy Quality: A Multiple Constituency Perspective," *Assuring business processes, proc. of the 6th annual security conference, Global Publishing, USA, Washington DC , April, pp 1-14*.

Maynard, S., Ruighaver, A., and Ahmad, A. 2011. "Stakeholders in Security Policy Development," *9th Australian Information Security Management Conference*, December , pp 182-188.

Maynard, S.B. 2010. "Strategic Information Security Policy Quality Assessment: A Multiple Constituency Perspective," PhD Thesis, *Department of Information Systems*, The University of Melbourne.

Neuman, W.L. 2006. "Social Research Methods: Qualitative and Quantitative Approaches," Allyn and Bacon.

Okoli, C., and Schabram, K. 2010. "A Guide to Conducting a Systematic Literature Review of Information Systems Research," *prouts: Working Papers on Information Systems.*.

Ølnes, J. 1994. "Development of Security Policies," *Computers & Security* (13:8), pp 628-636.

Oost, D., and Chew, E.K. 2012. "Investigating the Concept of Information Security Culture," in: *Strategic and Practical Approaches for Information Security Governance: Technologies and Applied Solutions*. IGI Global, pp 1-12.

Palmer, M.E., Robinson, C., Patilla, J.C., and Moser, E.P. 2001. "Information Security Policy Framework: Best Practices for Security Policy in the E-Commerce Age," *Information Systems Security* (10:2), May , pp 1-15.

Park, S., Ruighaver, A.B., Maynard, S.B., and Ahmad, A. 2012. "Towards Understanding Deterrence: Information Security Managers' Perspective," in: *Proceedings of the International Conference on IT Convergence and Security*. Suwon, Korea.

Patrick, D.H. 2002. "The Security Policy Life Cycle," in: *Information Security Management Handbook, Fourth Edition, Volume 4*. Auerbach Publications, pp 297-311.

Peltier, T.R. 2013. *Information Security Policies, Procedures, and Standards: Guidelines for Effective Information Security Management*. CRC Press.

Puhakainen, P., and Siponen, M. 2010. "Improving Employees' Compliance through Information Systems Security Training: An Action Research Study," *Mis Quarterly* (34:4), pp 757-778.

Ramachandran, S., Rao, C., Goles, T., and Dhillon, G. 2012. "Variations in Information Security Cultures across Professions: A Qualitative Study," *Communications of the Association for Information Systems (33:11) December pp 163-204.*

Rees, J., Bandyopadhyay, S., and Spafford, E.H. 2003. "PFIRS: A Policy Framework for Information Security," *Communications of the ACM* (46:7), pp 101-106.

Ruighaver, Maynard, S.B., and Chang, S. 2007. "Organisational Security Culture: Extending the End-User Perspective," *Computers & Security* (26:1), pp 56-62.

SANS Institute. 2001. "Security Policy Roadmap - Process for Creating Security Policies."

Siponen, M., Adam Mahmood, M., and Pahnila, S. 2014. "Employees' Adherence to Information Security Policies: An Exploratory Field Study," *Information & Management* (51:2), December , pp 217-224.

Siponen, M., Pahnila, S., and Mahmood, A. 2007. "Employees' Adherence to Information Security Policies: An Empirical Study," in: *New Approaches for Security, Privacy and Trust in Complex Environments*. Springer, pp 133-144.

Sommestad, T., Hallberg, J., Lundholm, K., and Bengtsson, J. 2014. "Variables Influencing Information Security Policy Compliance: A Systematic Review of Quantitative Studies," *Information Management & Computer Security* (22:1), pp 42-75.

Stahl, B.C., Doherty, N.F., and Shaw, M. 2012. "Information Security Policies in the Uk Healthcare Sector: A Critical Evaluation," *Information Systems Journal* (22:1), pp 77-94.

Vance, A., Siponen, M., and Pahnila, S. 2012. "Motivating Is Security Compliance: Insights from Habit and Protection Motivation Theory," *Information & Management* (49:3–4), pp 190-198.

Webb, J., Ahmad, A., Maynard, S.B., and Shanks, G. 2014. "A Situation Awareness Model for Information Security Risk Management," *Computers & Security* (44), pp 1-15.

Whitman, M.E. 2008. "Security Policy: From Design to Maintenance," in: *Information Security : Policy, Processes, and Practices. Advances in Management Information Systems.,* D.W. Straub, S.E. Goodman and R. Baskerville (eds.). London, England Armonk, New York : M.E. Sharpe, pp. 123-151.

Whitman, M.E., and Mattord, H.J. 2010. *Management of Information Security*. CengageBrain.

Whitman, M.E., Townsend, A.M., and Aalberts, R.J. 1999. "Considerations for an Effective Telecommunications-Use Policy," *Communications of the ACM* (42:6), pp 101-108.

Whitman, M.E., Townsend, A.M., and Aalberts, R.J. 2001. "Information Systems Security and the Need for Policy,").

Wood, C.C. 1995. "Writing Infosec Policies," *Computers & Security* (14:8), pp 667-674.

Wood, C.C. 2005. *Information Security Policies Made Easy : A Comprehensive Set of Information Security Policies*. Houston : InformationShield, c2005. Version 10.0.

Wood, C.C., and Lineman, D. 2009. *Information Security Policies Made Easy Version 11*. Information Shield, Inc.